\documentclass[onecolumn,preprintnumbers,amsmath,amssymb,10pt,a4paper,showkeys,superscriptaddress,nofootinbib]{revtex4}
\usepackage{geometry}
\usepackage{graphicx}
\usepackage{dcolumn}
\usepackage{bm}
\newcommand{\be}{\begin{equation}}
\newcommand{\ee}{\end{equation}}
\newcommand{\bd}{\begin{displaymath}}
\newcommand{\ed}{\end{displaymath}}
\newcommand{\BE}{\begin{eqnarray}}
\newcommand{\EE}{\end{eqnarray}}

\newcommand{\bq}{\ensuremath{\mathbf{q}}}
\newcommand{\bs}{\ensuremath{\mathbf{s}}}

\newcommand{\bx}{\ensuremath{\mathbf{x}}}
\newcommand{\by}{\ensuremath{\mathbf{y}}}

\newcommand{\avg}[1]{\left\langle{#1}\right\rangle}

\begin{document}

\preprint{}
\title{Fixation and escape times in stochastic game learning}


\author{John Realpe-Gomez}
\email{john.realpe@polito.it}
\affiliation{Politecnico di Torino, Corso Duca degli Abruzzi 24, 10129 Torino, Italy}

\author{Bartosz Szczesny}
\email{mmbs@leeds.ac.uk}
\affiliation{Theoretical Physics, School of Physics and Astronomy, The University of Manchester, Manchester M13 9PL, United Kingdom}
\affiliation{Department of Applied Mathematics, School of Mathematics, University of Leeds, Leeds LS2 9JT, United Kingdom}

\author{Luca Dall'Asta}
\email{luca.dallasta@polito.it}
\affiliation{Politecnico di Torino, Corso Duca degli Abruzzi 24, 10129 Torino, Italy}
\affiliation{The Abdus Salam International Centre for Theoretical Physics (ICTP), Strada Costiera 11, 34014 Trieste, Italy}

\author{Tobias Galla}
\email{tobias.galla@manchester.ac.uk}
\affiliation{Theoretical Physics, School of Physics and Astronomy, The University of Manchester, Manchester M13 9PL, United Kingdom}

\date{\today}

\begin{abstract}
Evolutionary dynamics in finite populations is known to fixate eventually in the absence of mutation. We here show that a similar phenomenon can be found in stochastic game dynamical batch learning, and investigate fixation in learning processes in a simple $2\times 2$ game, for two-player games with cyclic interaction, and in the context of the best-shot network game. The analogues of finite populations in evolution are here finite batches of observations between strategy updates. We study when and how such fixation can occur, and present results on the average time-to-fixation from numerical simulations. Simple cases are also amenable to analytical approaches and we provide estimates of the behaviour of so-called escape times as a function of the batch size. The differences and similarities with escape and fixation in evolutionary dynamics are discussed.
\end{abstract}

\maketitle

\section{Introduction}
Modern approaches to game theory have moved beyond the identification of equilibrium points of games \cite{Nash1950, Nash1951,Neumann1953}, and instead consider dynamical processes in populations of agents \cite{MaynardSmith1998, nowakbook,  Sigmund2010}, or the adaptation of a given set of agents to each other's actions \cite{fudenberg,Young2004, Camerer2003}. The study of populations of players is the focus of what is now called `evolutionary game theory' \cite{MaynardSmith1998, Vega2003,Gintis2000}. Within this field two approaches can broadly be distinguished. The more conventional one describes evolving populations by means of deterministic replicator equations, for textbooks see e.g. \cite{hofbauer}. The dynamical behaviour and attractors of these systems are studied with tools from the theory of nonlinear differential equations. Such formulations are valid formally only for infinite populations of agents, and systematically neglect stochastic effects in finite populations. The study of these random processes is at the centre of the second, more recent class of studies in evolutionary game theory, see for example \cite{traulsenreview} for a review. Crucial differences between the behaviour of finite and of infinite populations have been identified, for example finite systems may fixate at pure-strategy absorbing states even when the corresponding deterministic replicator equations have their attractors at mixed equilibria.  These stochastic processes are studied with a variety of different tools, including the master equation formalism, system-size expansions, backwards Fokker-Planck methods, and other concepts from statistical mechanics \cite{kampen,risken, gardiner}.

The purpose of the present paper is to parallel existing research on stochastic effects in evolutionary systems with studies of corresponding effects in stochastic learning dynamics. Learning is here related to, but different from evolution. Learning, or adaptation, is concerned with a fixed set of players who interact repeatedly in a given game, and who react to their opponent's actions by modifying their own strategic propensities \cite{fudenberg, Young2004}. These processes occur on much shorter time scales than evolutionary dynamics. Adaptation dynamics of the type we study here are of interest in two main different contexts. First, learning models provide mathematical descriptions of human or animal decision making and can be used to model the outcome of experiments in behavioural game theory and cognitive science \cite{Camerer2003}. The second main area in which models of adaptation are relevant is in machine learning and algorithmic game theory \cite{Nisan2007}. Here the interest is not in the modelling of human behaviour, but instead in the properties and design of algorithms with which to identify equilibrium points or solutions of optimisation problems. Understanding the dynamics of learning is of key importance in both of these applications. 

In learning there are no birth-death processes as in evolution, but instead dynamical updates of the agents' strategy profiles in time. Very little work exists on the systematic comparison of the effects of noise in evolution and in learning. Initial investigations \cite{galla} have shown, that similar to what is seen in evolutionary processes, the dynamics and attractors of stochastic learning can be quite different from that of deterministic adaptation processes. Up to now the analyses of fluctuations in learning are however limited to the identification of so-called quasi-cycles, also seen in evolution \cite{bladon,mobilia}. In the present paper we aim to establish further analogies between the two modelling approaches, and focus in particular on fixation effects \cite{antal, altrockinger}. Fixation here refers to processes by which dynamical systems reach absorbing states. In evolution these are typically points at the boundaries of strategy space, at which only one species (pure strategy) survives, and where all other strategies are extinct. In finite populations the elimination of species may happen by random drift, and in the absence of mutation a species is then never introduced again in the dynamics once all its representatives have been removed. The system thus fixates in an absorbing state.

In this paper we investigate the extent to which a similar removal of strategies may occur in multi-player learning. The analogue of extinction is here the convergence of a player's strategic propensities to a pure strategy.  The question we address is here when and how stochastic learning fixates. In particular we ask (i) under what circumstances convergence to pure, rather than mixed equilibra occurs in learning, (ii) if fixation occurs what are the corresponding extinction times, and (iii) given that extinction phenomena are well known in evolutionary systems, what are the differences and similarities with fixation in learning ? To answer these questions we consider several different types of games. After a general introduction to learning and the required definitions in Sec. \ref{sec:def} we first study simple two person games in Sec. \ref{sec:hd}. We then turn to games with cyclic interaction in Sec. \ref{sec:rps}, before we finally discuss a more intricate best-shot game \cite{galeotti,asta1,asta2} defined on regular random graphs (Sec. \ref{sec:net}). The final section summarises our results and discusses possible future work.

\section{Deterministic and stochastic learning}\label{sec:def}
\subsection{General definitions}
In this paper we will consider both two-player and multi-player games. Interaction will occur in learning processes, in which each player interacts only with a small number $p-1$ of other agents, in two-player games we will have $p=2$, for multi-player games one has $p>2$. Individual players will typically be labelled by indices $i\in\{1,\dots,M\}$, where $M$ stands for the total number of players in the model at hand. We will restrict the discussion to symmetric non-cooperative games. The variable $S$ will indicate the number of pure strategies available to each of the players. Following the standard game theoretic notation we will write $u(s,\bs_{-i})$ for the payoff player $i$ receives when playing pure strategy $s\in\{1,\dots,S\}$, and when her opponents play actions $\bs_{-i}\in\{1,\dots,S\}^{p-1}$. This paper focuses only on symmetric games so that $u(\cdot,\cdot)$ is identical for all players, and carries no explicit dependence on $i$. We will use the notation $\bx_i$ for player $i$'s mixed strategy, i.e. we have $\bx_i=(x_{i,1},\dots,x_{i,S})$ with $\sum_s x_{i,s}=1$ for all $i$. The component $x_{i,s}$ indicates the frequency with which player $i\in\{1,\dots,M\}$ plays pure strategy $s\in\{1,\dots,S\}$.

\subsection{Learning}

We will here focus on a re-inforcement type learning model, and assume that each player keeps a score valuation of each of his/her pure strategies, these are a measure of the (perceived) relative performance of the pure actions in the past, and indicate the propensity of playing any particular pure action. Discarding memory-loss, the valuation $q_{i,s}(t)$ player $i$ has for pure strategy $s$ is the cumulative payoff $i$ would have received in all past rounds up to time $t$, given his opponent's actions, and had $i$ always played pure strategy $s$ up to time $t$.  This will be detailed further below. Following \cite{Camerer2003,Ho2007,satopre,satopnas} we will assume that given the score valuation vector $\bq_i(t)=(q_{i,1}(t),\dots,q_{i,S}(t))$ player $i$ chooses each of the pure strategies according to a logit rule, i.e. that the probabilities of playing the different pure strategies depend on the score valuations via the following relations:
\be x_{i,s}(t)=\frac{e^{\beta q_{i,s}(t)}}{\sum_{s'}e^{\beta q_{i,s'}(t)}}\label{eq:probx}. 
\ee 
The variable $\beta$ is here a model parameter, and describes a learning rate or intensity of selection. For $\beta\to \infty$ the players strictly choose the pure action with highest propensity. For $\beta=0$ they play at random.

A learning dynamics is then a description governing the evolution of the $q_{i,s}(t)$ in time. We will here mostly focus on a re-inforcement learning rule of the form
\BE
q_{i,s}(t+1)&=&(1-\alpha) q_{i,s}(t)+u(s,\bs_{-i}(t)).\label{eq:qrupdate}
\EE
The interpretation of these update rules is understood best by first considering the case $\alpha=0$: in this case the increment of $q_{i,s}$ between time-steps $t$ and $t+1$ is the payoff player $i$ would have received had they played pure strategy $s$, and given their opponents actions $\bs_{-i}(t)$. For $\alpha=0$ the variable $q_{i,s}(t)$ is thus the total payoff player $i$ would have received given their opponent's play had $i$ always played action $s$. A non-zero value of $\alpha$  accounts for exponential discounting over time, or equivalently for a possible memory loss. For $\alpha>0$ the outcomes of the game in the distant past have a lesser effect on the valuation $q_{i,s}(t)$ than the more recent rounds of the game.  

The process defined by Eq. (\ref{eq:qrupdate}) is inherently stochastic, given that all players choose their pure actions according to the probabilistic rules of Eq. (\ref{eq:probx}). A deterministic limit has been considered in \cite{satopre,satopnas,satophysica}, and can be formulated as
\BE
q_{i,s}(\tau+1)&=&(1-\alpha) q_{i,s}(\tau)+\sum_{\bs_{-i}} u(s,\bs_{-i})\bx_{-i,\bs_{-i}}(\tau),\label{eq:detqrupdate}
\EE
where $\bx_{-i,\bs_{-i}}$ stands for the probability of action $\bs_{-i}\in\{1,\dots,S\}^{p-1}$ being played by $i$'s opponents, i.e. we have $\bx_{-i,\bs_{-i}}=\prod_{j\neq i} x_{j,s_j}$. Taking into account Eqs. (\ref{eq:probx}) one can then write the update rule solely in terms of $\bx$ and $\by$ and finds the following map \cite{satopre}
\BE
x_{i,s}(\tau+1)&=&\frac{x_{i,s}(\tau)^{1-\alpha}e^{\beta \sum_{\bs_{-i}} u(s,\bs_{-i})\bx_{-i,\bs_{-i}}(\tau)}}{\sum_{s'} x_{i,s'}(\tau)^{1-\alpha}e^{\beta \sum_{\bs_{-i}} u(s',\bs_{-i})\bx_{-i,\bs_{-i}}(\tau)}}.\label{eq:detmap}
\EE
To interpolate between the stochastic process defined by Eqs. (\ref{eq:probx},\ref{eq:qrupdate}) and the deterministic limit of Eq. (\ref{eq:detqrupdate}) we will consider a batch learning process, in which players update their score valuations only once every $N$ rounds of the game, and keep them constant inbetween. Specifically, we will assume
\BE
q_{i,s}(t+N)&=&(1-\alpha) q_{i,s}(t)+\frac{1}{N}\sum_{t'=t}^{t+N-1}u(s,\bs_{-i}(t')),\label{eq:batchqrupdate}
\EE
and $q_{i,s}(t+\ell)=q_{i,s}(t)$ for all $\ell=1,2,\dots,N-1$ . We will refer to $N$ as the batch size of the learning process. The batch process at $N>1$ (but finite) is here mostly a theoretical vehicle which allows one to understand the dynamics of learning. Real-world adaptation presumably operates close to the limit $N=1$, nevertheless some of the existing work has focused on deterministic learning ($N\to\infty$). Our work tries to address the gap between these two extreme cases, and to establish in a systematic manner the stochastic effects affecting the dynamics at finite batch sizes. The case $N=1$ can be understood as a special limiting case. Previous work has shown that approaches taken based on a systematic expansion in $1/\sqrt{N}$ can give good results even for small batch sizes \cite{galla}.

\subsection{Sato-Crutchfield dynamics in continuous time}
In order to make contact with deterministic descriptions of evolutionary systems it is helpful to consider the continuous-time limit of the deterministic learning process, Eq. (\ref{eq:detmap}). Assuming the validity of such a limit for small intensity of selection, $\beta\ll 1$, and following \cite{satopre,satophysica} one finds 
\BE
\dot x_{i,s}&=&\beta x_{i,s}\left(\sum_{\bs_{-i}}u(s,\bs_{-i})\bx_{-i,\bs_{-i}}-\sum_{s',\bs_{-i}} x_{i,s'} u(s',\bs_{-i})\bx_{-i}\right)-\alpha x_{i,s}\left(\log x_{i,s}-\sum_{s'} x_{i,s'}\log x_{i,s'}\right).
\label{eq:sato}
\EE
For $\alpha=0$ this reduces to a set of multi-population replicator equations, a signature of the close connection between evolutionary processes and adaptive learning.

\section{Two-player Hawk-Dove game}\label{sec:hd}

\subsection{Definition and replicator flow}
\begin{figure}
\centerline{\includegraphics[width=0.4\textwidth]{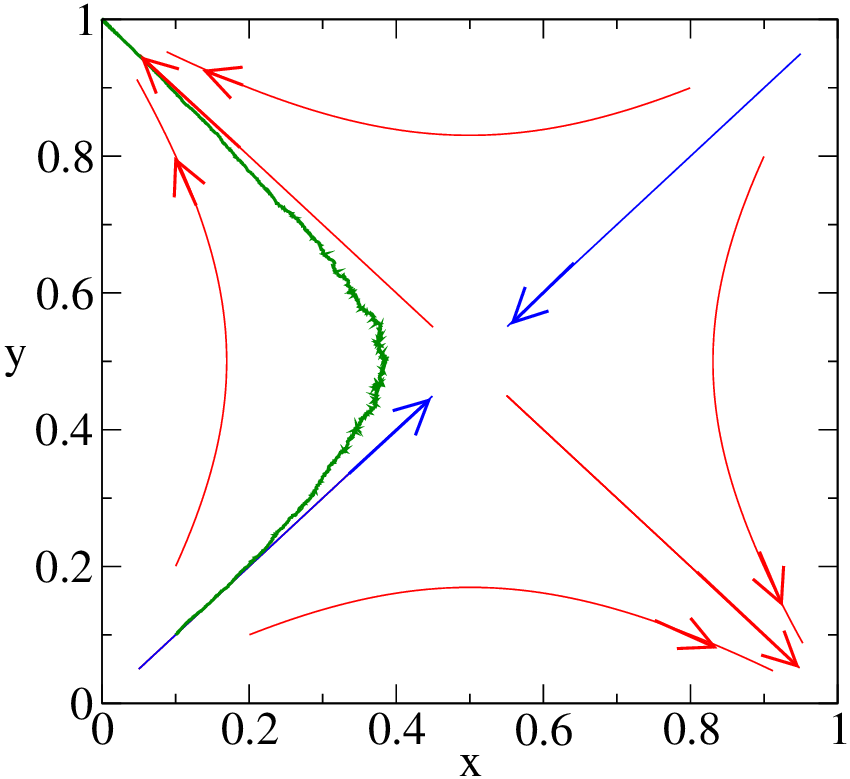}}
\caption{\label{fig:hdflow}(Colour on-line) Deterministic flow of the two-population replicator dynamics for the Hawk-Dove game (arrows). The noisy green line shows the trajectory of one single realisation of the learning dynamics at batch size $N=5$, started at $x(0)=y(0)=0.1$, and with parameters $\alpha=0, \beta=0.01$.}
\end{figure}
We will first consider the case of a symmetric $2\times 2$ game, the so-called Hawk-Dove game (also referred to as the coexistence game or the anti-coordination game) defined by the payoff matrix
\be
A=\left(\begin{array}{cc}  (b-c)/2 & b \\ 0 & b/2 \end{array}\right), \label{eq:hdpayoff}
\ee
where we set $b=1$ and $c=2$. We will label the elements of the payoff matrix by $a_{s,s'}$, where $s$ and $s'$ can each take one of two values, representing the pure strategies of this game. In the learning process two players interact repeatedly, the strategy of each player is fully characterized by the probability of playing `Hawk'. We will denote these probabilities by $x(t)$ for the first player, and by $y(t)$ for player 2. In the absence of memory loss, and taking the continuous-time limit of Eqs. (\ref{eq:detmap}) we obtain the two-population replicator dynamics
\BE
\dot x&=&x(1-x)\left(\frac{1}{2}-y\right) \nonumber \\
\dot y&=&y(1-y)\left(\frac{1}{2}-x\right).
\EE
These equations are obtained from setting $\beta=1, \alpha=0$ in the above Sato-Crutchfield equations (\ref{eq:sato}) and upon using $u(s_i,s_j)=a_{s_i,s_j}$ with the above payoff matrix (\ref{eq:hdpayoff}). It is straightforward to work out the  corresponding deterministic flow, we illustrate it for completeness in Fig. \ref{fig:hdflow}. The replicator dynamics has one reactive fixed point at $(x^*,y^*)=(1/2,1/2)$, and two pure strategy fixed points at $(0,1)$ and $(1,0)$. These fixed points at the boundary of strategy space are stable attractors, the central fixed point is a saddle with one stable and one unstable eigendirection. The stable eigenvector points along the diagonal, and restricting the dynamics to this direction (i.e. setting $x=y$) hence yields a stable flow towards the central fixed point. The single-population replicator equation 
\be
\dot x=x(1-x)(1/2-x)
\ee
therefore converges to $x=1/2$, provided non-extremal initial conditions ($x(0)\neq 0, x(0)\neq 1$) are chosen. The two-population system will generally fixate at one of the corner attractors for generic initial conditions $x(0)\neq y(0)$, only in the restricted case $x(0)=y(0)$ is the symmetry between the players preserved and dynamics converges to the mixed fixed point.

\subsection{Fixation in stochastic learning}
\begin{figure}
\centerline{\includegraphics[width=0.4\textwidth]{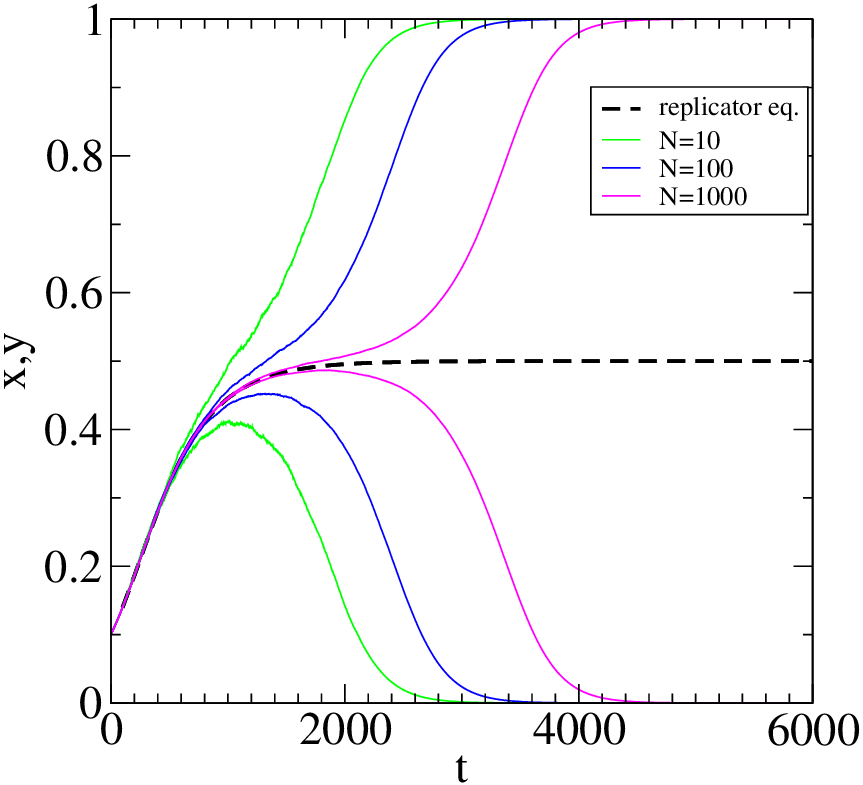}~~~~~~~~~\includegraphics[width=0.4\textwidth]{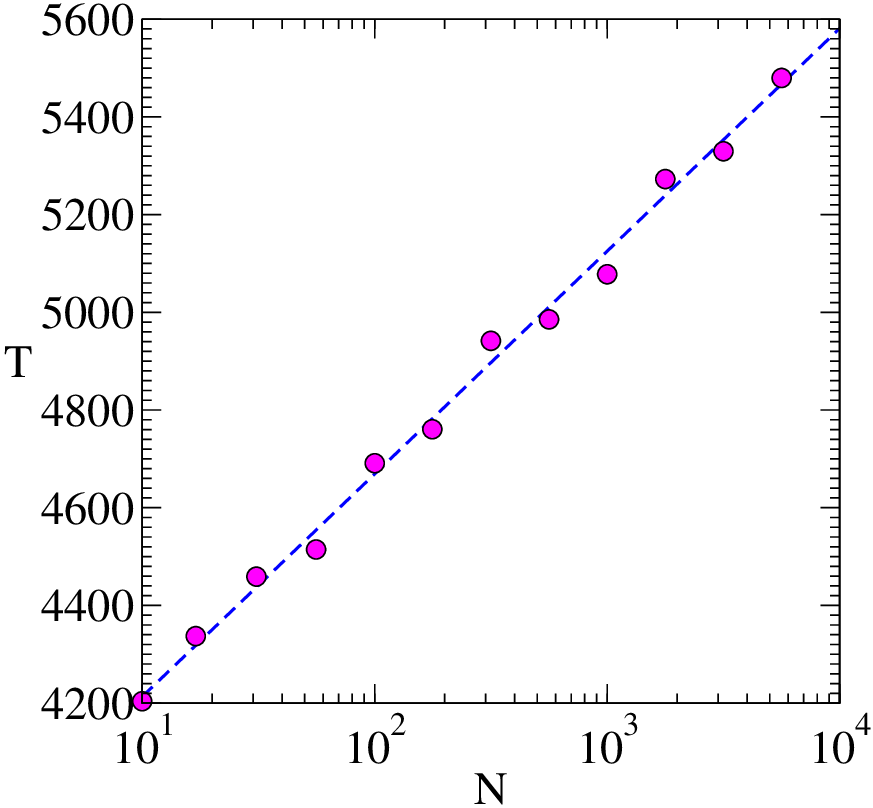}}
\caption{(Colour on-line) \label{fig:fix_hd} Fixation in stochastic learning of the Hawk-Dove game ($\alpha=0$). Left panel shows individual trajectories for different batch sizes $N$, started at $x(0)=y(0)=0.1$. Each pair of curves shows $x(t)$ and $y(t)$ for a single run. The black dashed line indicates the evolution of the deterministic replicator equations. Parameters are $\alpha=0$ and $\beta=0.01$. The right panel shows the mean extinction time as a function of the batch size (simulations started from $x(0)=y(0)=1/2$, extinction as defined in the main text). Symbols are from simulations (initial conditions $x(0)=y(0)=0.1$, averaged over $100$ independent runs, $\alpha=0,\beta=0.01$). The dashed line is a fit to a logarithmic dependence $T=c_1+c_2\ln N$.}
\end{figure}
We will first address learning dynamics in absence of memory-loss ($\alpha=0$), the effects of exponential discounting are described in Sec. \ref{sec:alpha}. Numerical simulations show that stochastic learning without memory-loss will generally fixate in one of the two corners, $(1,0)$ or $(0,1)$ of strategy space, a typical trajectory generated by the learning dynamics at finite batch size is shown in Fig. \ref{fig:hdflow}. This is further illustrated in the left panel of Fig. \ref{fig:fix_hd}, where we show the evolution of $x(t)$ and $y(t)$ in stochastic learning at different batch sizes $N$. The dynamics are here started from $x(0)=y(0)$, and will initially follow the replicator flow closely, and approach, but not reach the replicator fixed point at $(1/2,1/2)$. Fluctuations, which will invariably occur at any finite batch size, break the symmetry between $x$ and $y$ however, and the system will generally drift off the diagonal $x=y$ relatively quickly. While the stable eigenvalue of the central fixed point still exerts some attraction to the centre, the unstable direction will eventually take over, and draw the learning process to $(1,0)$ or $(0,1)$. Which one of these corners is reached is purely random, and determined by the nature of sampling errors in the adaptation process. Large batch sizes here reduce the amount of noise in the dynamics, and the system hence follows the deterministic flow longer at large $N$ than at smaller batches, as illustrated in the left panel of Fig. \ref{fig:fix_hd}. In the right panel of the figure we have measured the time-to-fixation $T$ more systematically. Specifically we consider the system to be fixated once $x(t)$ and $y(t)$ have each approached the values $0$ or $1$ up to an accuracy $\vartheta$, with $\vartheta>0$ a small threshold ($\vartheta=10^{-5}$ in Fig. \ref{fig:fix_hd}). Once this condition is met each player plays essentially a pure strategy, i.e. the system is close to one of the corners of strategy space  up to deviations smaller than $\vartheta$. We find logarithmic behaviour of the so-defined fixation time, i.e. $T\sim \ln N$. This is consistent with observations in one-dimensional evolutionary co-ordination games, with one central unstable fixed point \cite{nowakbook,traulsenreview}.

It is generally very hard to compute the time-to-fixation of stochastic processes analytically, this applies both to learning processes and to evolutionary dynamics. In the latter, general analytical results have been obtained only for one-population models \cite{nowakbook,traulsenreview}. One major complication is here the fact that the dynamics in most other cases has at least two degrees of freedom, impeding a full analytical solution.

\begin{figure}[t]
\vspace{3em}
\centerline{\includegraphics[width=0.5\textwidth]{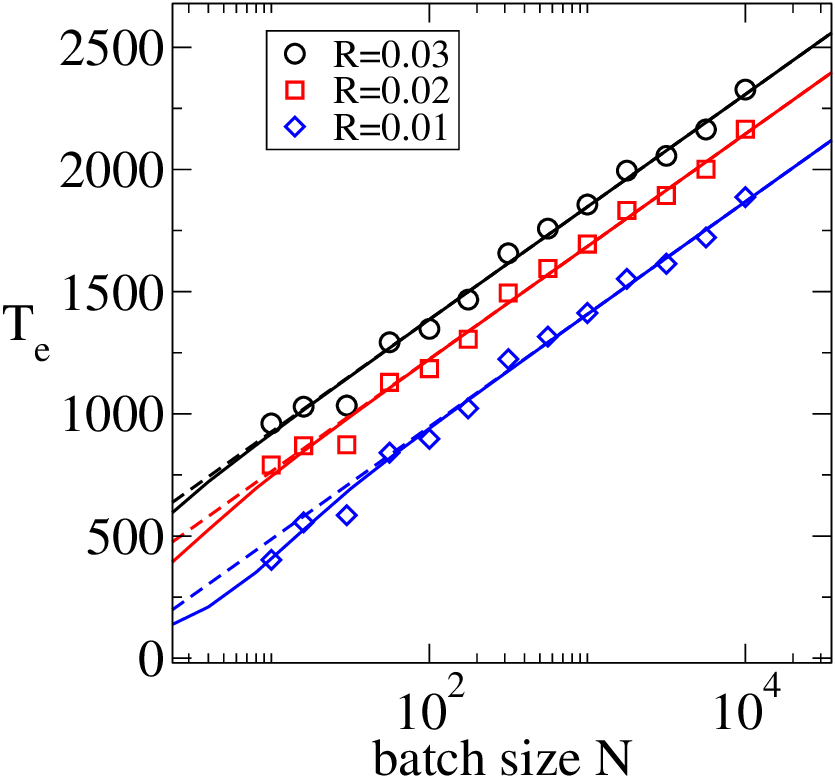}}
\caption{\label{fig:hdlearning}(Colour on-line) Two-player learning dynamics in the Hawk-Dove game. The figure shows the escape time from a ball about the central fixed point (see text for details). Symbols are results from numerical simulations ($\beta=0.01$, averaged over $100$ samples), solid lines show the theoretical estimates of Eq. (\ref{eq:genescape}), dashed lines the approximation of Eq. (\ref{eq:lgtz}). The escape time scales logarithmically in the system size $N$, in-line with the existence of an unstable eigendirection of the limiting deterministic dynamics.}
\end{figure}

Partial analytical results for game dynamical learning can however be obtained for what we will refer to as `escape times' in the following, see also \cite{mobilia} for studies of escape times in cyclic evolutionary games. For a given (finite) batch size $N$ we here start the learning dynamics at the deterministic fixed point $x(0)=y(0)=1/2$, and run the stochastic dynamics until the system reaches a given distance $R$ from this fixed point. More precisely we define the escape time $T(R)$ as the time at which the variable $|x(t)-y(t)|/\sqrt{2}$ first exceeds the value $R$. This measure of distance was chosen for analytical convenience, as it will become clear below. Results from simulations are shown in Fig. \ref{fig:hdlearning}.
Analytical predictions of the escape time for small values of $R$ are possible within a linear approximation about the central fixed point. Using the methods detailed in \cite{galla,galla2} we find that in the limit $\beta\ll 1$ and for large but finite batch size $N$ the two-player learning dynamics can be described by the following Langevin dynamics 
\BE
\dot {\widetilde x} &=& J_{11} \widetilde x + J_{12}\widetilde y + \xi \nonumber  \\
\dot {\widetilde y} &=& J_{21} \widetilde x + J_{22}\widetilde y + \eta,
\EE
where the matrix $\mathbb{J}$ is the Jacobian of the continuous-time learning dynamics (equivalent to the replicator equations for the case $\alpha=0$ we are considering here), specifically we have  $J_{11}=J_{22}=0, J_{12}=J_{21}=-\beta/4$ at vanishing memory loss\footnote{For general $\alpha$ one has $J_{11}=J_{22}=-\alpha$.}. We have here introduced $\widetilde x(t)=x(t)-1/2$ and $\widetilde y(t)=y(t)-1/2$, and address only the stationary regime in which deterministic learning has assumed its fixed point. The variables $\widetilde x$ and $\widetilde y$ describe fluctuations about this fixed point, $\xi$ and $\eta$ represent Gaussian white noise, with variances and correlations given by
\be
\avg{\xi(t)\xi(t)}=\avg{\eta(t)\eta(t)}=\frac{\beta^2}{64N}, ~~~ \avg{\xi(t)\eta(t)}=0.
\ee  
Given that $(1,-1)$ is an eigenvector of the above Jacobian (with an eigenvalue of $\beta/4$) we then have
\be
\dot d=\frac{\beta}{4}d+\zeta,
\ee
where $d(t)=(x(t)-y(t))/\sqrt{2}$, and where $\avg{\zeta(t)\zeta(t')}=\sigma^2N^{-1}\delta(t-t')$, with $\sigma^2=\beta^2/64$. Using results for escape times of general Langevin processes of the form $\dot x = \lambda x+\eta$ with $\avg{\eta(t)\eta(t')}=(\sigma^2/N)\delta(t-t')$ (see appendix) we then obtain the following prediction for the escape time
\be\label{eq:genescape}
T(R)=2\int_0^{w_N} \mathrm{d}w ~ \frac{e^{-w}}{\lambda w}~ \sinh\left(w-\frac{w^2}{w_N}\right),
\ee
with $w_N = N\lambda R^2/\sigma^2$. In our specific example we have $\lambda=\beta/4$ and $\sigma^2=\beta^2/64$. As seen in Fig. \ref{fig:hdlearning} this compares well with numerical simulations.

\subsection{Comparison with evolutionary dynamics in finite populations}
We have already indicated that the behaviour of the stochastic learning dynamics is, to an extent, similar to evolutionary processes. To quantify this further we investigate both one-population and two-population stochastic evolutionary processes in this section. 

\subsubsection{Two-population dynamics}
Specifically we consider two populations, each composed of $N$ players. Each of these players will either be a Hawk or a Dove, we denote the number of Hawks in the first population by $i$, and the number of Hawks in the second population by $j$ respectively. The corresponding numbers of Doves are then $N-i$ and $N-j$ in the two populations. Players in the first population only play against players of the second population, and vice versa. The fitness of a Hawk and Dove players in the first population are then for example given by
\be\label{eq:fit}
f_{1,H}=1-\frac{3}{2}\frac{j}{N}, ~~ f_{1,D}=\frac{1}{2}-\frac{1}{2}\frac{j}{N},
\ee
and similar definitions hold for individuals in the second population. In order to specify a microscopic dynamics we use the so-called `local update rule', sometimes also referred to as the `pairwise comparison process' \cite{traulsenreview}. A player of the `Dove' type is converted into `Hawk' player with a rate proportional to $1+f_{m,H}-f_{m,D}$, where $m=1,2$ labels the two populations. Similarly conversions of Hawk players into Dove players occur with a rate proportional to $1+f_{m,D}-f_{m,H}$. Specifically, we will use the following transition rates
\BE
&&T_1^+ = \frac{1}{2} \frac{i}{N} \frac{N-i}{N} \left(\frac{3}{2} - \frac{j}{N} \right), ~~~
T_2^{+} = \frac{1}{2} \frac{j}{N} \frac{N-j}{N} \left(\frac{3}{2} - \frac{i}{N} \right)  \nonumber \\
&&T_1^{-} = \frac{1}{2} \frac{i}{N} \frac{N-i}{N} \left(\frac{1}{2} + \frac{j}{N} \right)  , ~~~
T_2^{-} = \frac{1}{2} \frac{j}{N} \frac{N-j}{N} \left(\frac{1}{2} + \frac{i}{N} \right).\label{eq:rates}
\EE
The factors of the form $(i(N-i))/N^2$ here indicate that two players of different types need to be drawn from any one population in order for an interaction to occur. It is here important to stress that reproduction and selection occurs {\em within} the separate populations, i.e. at no point is an individual of one population converted into a member of the other. Interaction between the population occurs via Eq. (\ref{eq:fit}), i.e. the fitness of members of population one depends on the composition of population two and vice versa.

\begin{figure}
\vspace{3em}
\centerline{\includegraphics[width=0.5\textwidth]{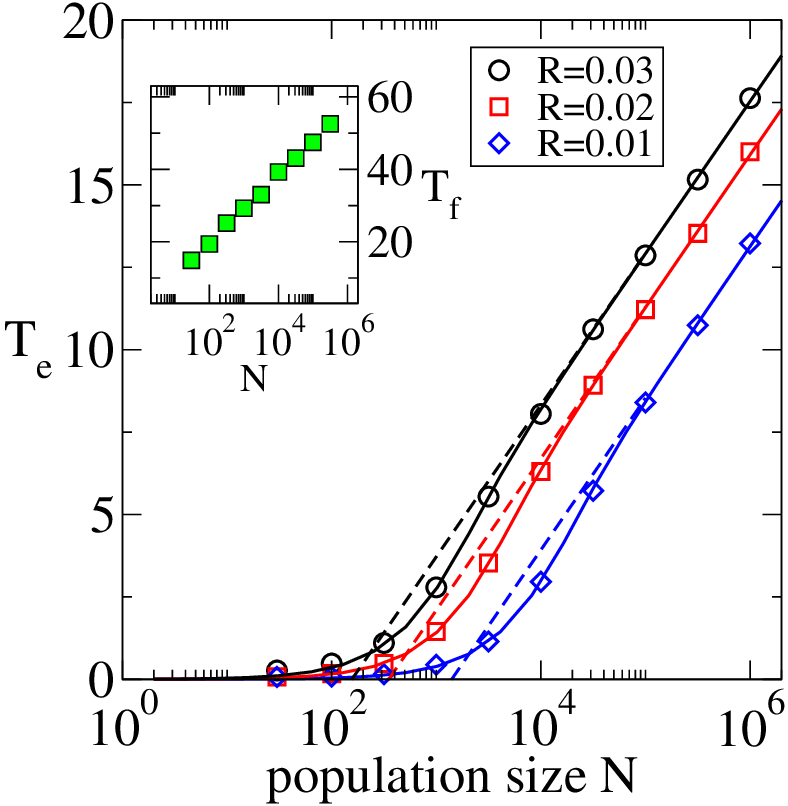}}
\caption{\label{fig:2pophd}(Colour on-line) Two-population evolutionary dynamics in the Hawk-Dove game: Main panel shows the escape time from a region about the central fixed point (see text for details). Symbols are results from numerical simulations (averaged over $1000$ samples), solid lines show the theoretical estimates of Eq. (\ref{eq:genescape}), dashed lines the approximation of Eq. (\ref{eq:lgtz}). The escape time scales logarithmically in the system size $N$, in-line with the existence of an unstable eigendirection of the limiting deterministic dynamics. The inset shows the fixation time as a function of the system size (average over $100$ runs). }
\end{figure}

In the deteministic limit ($N\to\infty$) one recovers the two-population replicator dynamics
\be\label{eq:twopoprepl}
\dot x = T_1^{+,\infty}-T_1^{-,\infty}=\frac{1}{2}x(1-x)(1-2y), ~~ \dot y = T_2^{+,\infty}-T_2^{-,\infty}=\frac{1}{2}y(1-y)(1-2x),
\ee
where we have used the replacements $\frac{i}{N}\to x$ and $\frac{j}{N}\to y$ in Eqs. (\ref{eq:rates}) to obtain the $T_m^{\pm,\infty}$. The deterministic flow of these replicator equations is the one indicated in Fig. \ref{fig:hdflow}, in particular the central fixed point has one stable and one unstable eigendirection. Fixation of the stochastic evolutionary dynamics can occur at any of the four corners of strategy space. We show results for the average time-to-fixation in the inset of Fig. \ref{fig:2pophd}, the fixation time depends logarithmically on the system size $N$.
As in the learning dynamics, an analytical calculation of the fixation time is very difficult. Estimates for the escape times can however be obtained within a system-size expansion about the fixed point of the deterministic replicator equations. Following standard methods based on the so-called `van Kampen expansion' in the inverse system size \cite{kampen} one finds
\BE
\dot {\widetilde x} &=& -\frac{1}{4}\widetilde y + \xi(t) \nonumber \\
\dot {\widetilde y} &=& -\frac{1}{4}\widetilde x + \eta(t)
\EE 
where $\widetilde x(t)=\frac{i}{N}-\frac{1}{2}$ and  $\widetilde y(t)=\frac{j}{N}-\frac{1}{2}$. As before $\xi(t)$ and $\eta(t)$ describe Gaussian noise, from the van Kampen expansion one finds $\avg{\xi(t)\eta(t')}=0$ as well as
\be
\avg{\xi(t)\xi(t')}=\avg{\eta(t)\eta(t')}=\frac{1}{4N}\delta(t-t').
\ee
This translates into a Langevin equation
\be
\dot d = \frac{1}{4}d+\zeta
\ee
for the variable $d(t)=(x(t)-y(t))/\sqrt{2}$, with $\avg{\zeta(t)\zeta(t')}=\frac{1}{4N}\delta(t-t')$. A theoretical prediction for the escape time can hence be found using the values $\lambda=\sigma^2=1/4$ in Eq. (\ref{eq:genescape}). Results are tested against simulations and confirmed successfully in Fig. \ref{fig:2pophd}.

\subsubsection{One-population dynamics}
In the one-population model one considers a single population of $N$ individuals, each of whom can either be a Hawk or a Dove player. The state of the system is hence characterized by a single integer, the number $i$ of Hawks. Transition rates of the local process read
\be\label{eq:onepoptrans}
T^+ = \frac{1}{2} \frac{i}{N} \frac{N-i}{N} \left(\frac{3}{2} - \frac{i}{N} \right), ~~~
T^{-} = \frac{1}{2} \frac{i}{N} \frac{N-i}{N} \left(\frac{1}{2} + \frac{i}{N} \right).  
\ee
The analysis of this model is not new as such, the study of single-population dynamics of $2\times 2$ games is in fact standard, see for example \cite{traulsenreview, altrockinger}. We here present results mainly for completeness and in order to contrast them with the above two-population case.

\begin{figure}
\vspace{3em}
\centerline{\includegraphics[width=0.5\textwidth]{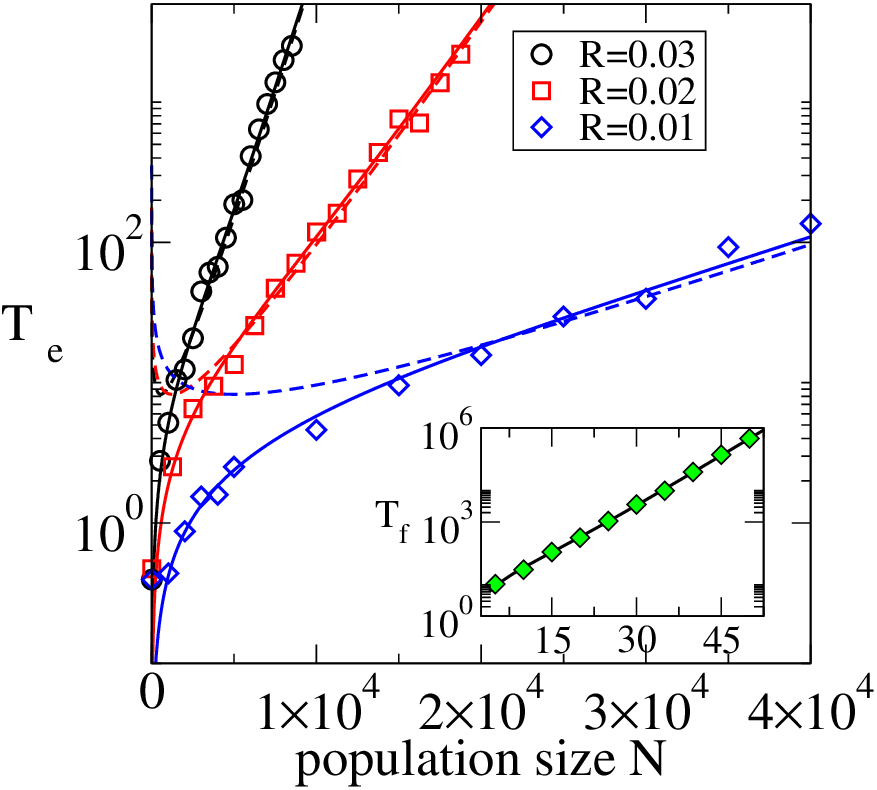}}
\caption{\label{fig:1pophd}(Colour on-line) One-population evolutionary dynamics in the Hawk-Dove game: main panel shows the escape time from the central fixed point (defined as the point in time at which the quantity $|\widetilde x(t)|$, with $\widetilde x(t)$ as defined in the text, first exceeds the value $R$). Symbols are results from numerical simulations (averaged over $20-50$ samples), solid lines show the theoretical results of Eq. (\ref{eq:genescape}), the dashed lines the asymptotic approximation of Eq. (\ref{eq:lltz}). The escape time scales exponentially in the system size $N$, in-line with the existence of a stable eigendirection of the limiting deterministic dynamics.The inset shows the fixation time as a function of the system size, symbols are from simulations (average over $100$ runs), the solid line from an analytical calculation based on the methods and expressions detailed in \cite{traulsenreview}. }
\end{figure}

In the deteministic limit ($N\to\infty$) the following replicator equation is obtained from Eq. (\ref{eq:onepoptrans}):
\be
\dot x = T^{+,\infty}-T^{-,\infty}=\frac{1}{2}x(1-x)(1-2x).
\ee
This corresponds to restricting the two-population replicator equations (\ref{eq:twopoprepl}) to the subspace in which $x(t)=y(t)$. In order to explore stochastic corrections to this limiting behaviour in next-to-leading order we again carry out the system-size expansion. As before we do not report the detailed mathematics, which is tedious, but standard. Defining $\widetilde x(t)=i/N-1/2$ one finds
\be
\dot {\widetilde x}(t)=-\frac{1}{4}\widetilde x(t)+\eta(t),
\ee
where $\avg{\eta(t)\eta(t')}=\frac{1}{4N}\delta(t-t')$. Setting $\lambda=-1/4$ and $\sigma^2=1/4$ in Eq. (\ref{eq:genescape}) we obtain semi-analytical predictions for the escape time. These results are compared with simulations in Fig. \ref{fig:1pophd}. As seen in the figure the escape time no longer scales logarithmically in the system size as it was the case in the two-population model, but instead the escape is now exponentially slow in the asymptotic limit of large $N$. We have also measured the actual fixation time (see inset of Fig. \ref{fig:1pophd}). Fixation times scale exponentially in the system size. Analytical results can here be obtained based on the methods described for example in \cite{traulsenreview}). For completeness we show the results of these calculations in the inset of Fig. \ref{fig:1pophd}.

\subsection{Effects of memory-loss in two-player learning}\label{sec:alpha}
\begin{figure}
\vspace{3em}
\centerline{\includegraphics[width=0.75\textwidth]{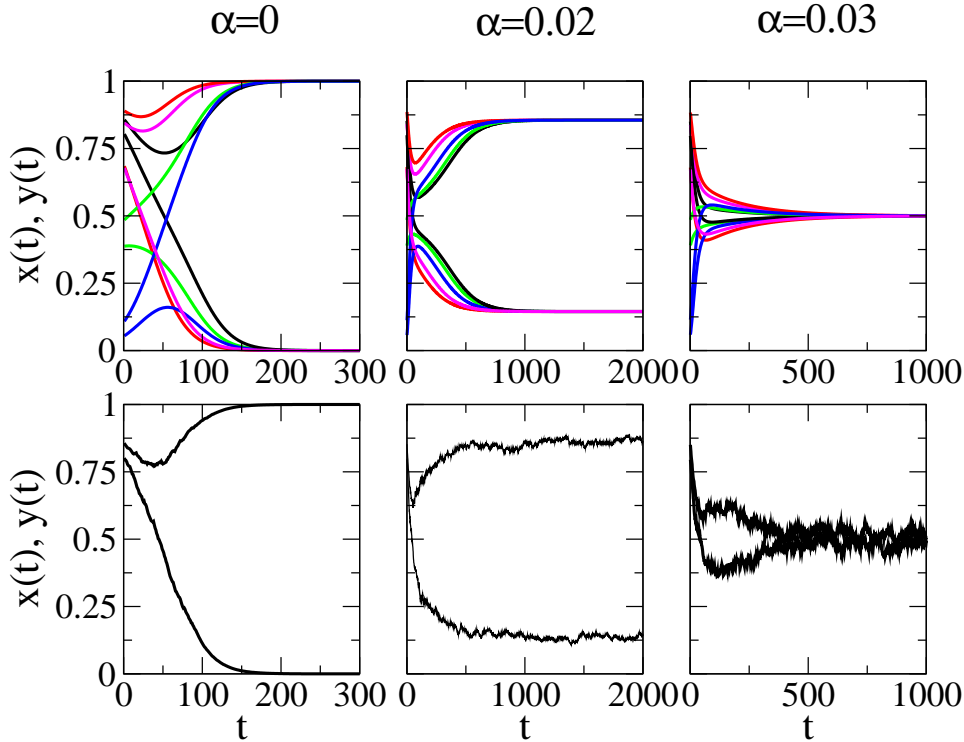}}
\caption{\label{fig:hdalpha} (Colour on-line) Effects of memory-loss on deterministic and stochastic learning in the Hawk-Dove game. The upper row shows the outcome of the deterministic dynamics at $\beta=0.1$ for different values of the memory-loss parameter $\alpha$. In each panel we show five trajectories ($x(t),y(t)$) obtained from five independent random initial conditions. Lower row: single runs $(x(t),y(t))$ of stochastic learning, started from random initial condition. The critical value $\alpha_c$ separating the regime of a stable central fixed point ($\alpha>\alpha_c$) from an unstable regime ($\alpha<\alpha_c)$ is given by $\alpha_c=\beta/4=0.025$.}
\end{figure}

Unlike in evolutionary dynamics, where fixation can occur in absence of mutation purely by random drift, fixation in stochastic game learning is strictly tied to the convergence of the limiting deterministic learning process to pure strategy equilibria. In order to demonstrate this we will extend the analysis to non-zero memory-loss rates $\alpha>0$ in the following.  Deterministic learning of the Hawk-Dove game in discrete time is then described by the two-dimensional map given by Eq. (\ref{eq:detmap}) (with the appropriate substitutions for the payoff structure). The point $(x^*,y^*)=(1/2,1/2)$ is a fixed point and the relevant eigenvalues are identified as $\lambda=(1-\alpha)\pm \beta/4$. Assuming $0\leq \alpha<1$ (and $\alpha<2-\beta/4$) we therefore find that the central fixed point is stable whenever $\alpha>\alpha_c=\beta/4$. In order to characterise the outcome of learning we have to distinguish between three different types of behaviour\footnote{In the context of this paper these are mostly an empirical observations in simulations and from numerical iteration of the deterministic learning process.}: 
\begin{enumerate}
\item[(1)] For $\alpha=0$ the central fixed point is not a stable attractor of the deterministic learning process. In this regime $(1,1)$ is still a stable eigendirection, so deterministic learning will converge to $(1/2,1/2)$ provided it is started from symmetric initial conditions ($x(0)=y(0)$). For generic initial conditions this symmetry is broken however, and the dynamics is observed to approach either $(1,0)$ or $(0,1)$ asymptotically. Noise in learning has a similar symmetry-breaking effect, and will drive the dynamics to one of the pure strategy attractors.
\item[(2)] For $\alpha>0$, but $\alpha<\alpha_c$ the central fixed point is again not a stable attractor of the learning dynamics, and deterministic learning will converge to $(1/2,1/2)$ only if started from symmetric initial conditions. For asymmetric initial conditions the dynamics will approach an asymmetric fixed point ($x^*\neq y^*$), which is generally not a pure strategy for $\alpha>0$. With noise learning fluctuates around this symmetry-broken attractor. Memory-loss in learning thus acts similar to mutation in evolutionary dynamics, and impedes absorption at the boundaries.
\item[(3)] For $\alpha>\alpha_c$ deterministic learning converges to $(1/2,1/2)$ even for non-symmetric initial conditions. In this case there is no fixation, the dynamics of stochastic learning will fluctuate around the mixed-strategy equilibrium asymptotically.
\end{enumerate}
This behaviour is illustrated further in Fig. \ref{fig:hdalpha}.

\section{Escape rates in cyclic games}\label{sec:rps}
We now consider a two-player discrete-time learning dynamics in the rock-paper-scissors game (RPS). Detailed analyses of evolutionary processes in this cyclic game can for example be found in \cite{mobilia}. We here focus on learning, and first concentrate on the deterministic limit. Specifically, using the deterministic limit  of Eqs. (\ref{eq:detmap}) we have the following map
\BE
x_s(t+1)=\frac{x_s(t)^{1-\alpha}e^{\beta R_s(t)}}{\sum_{s'} x_{s'}(t)^{1-\alpha}e^{\beta R_{s'}(t)}}, ~~~~ y_s(t+1)=\frac{y_s(t)^{1-\alpha}e^{\beta S_s(t)}}{\sum_{s'} y_{s'}(t)^{1-\alpha}e^{\beta S_{s'}(t)}}
\EE
where
\BE
R_s(t)=\sum_{s'} a_{ss'} y_{s'}(t), ~~~~~S_s(t)=\sum_{s'} a_{ss'} x_{s'}(t),
\EE
and where $A=(a_{ss'})$ ($s,s'\in\{R,P,S\}$) the standard RPS payoff matrix, i.e. 
\be
A=\left(\begin{array}{ccc} 0 &-1 &1 \\ 1 & 0 & -1 \\ -1 & 1 & 0\end{array}\right).
\ee
Due to overall normalisation, $\sum_sx_s=\sum_{s}y_{s}=1$, the above map defines a $4$-dimensional dynamical system. The mixed strategy point $x_s=y_s=1/3$ for all $s$ is always a fixed point, and the corresponding $4\times 4$ Jacobian is easily computed. One finds the following eigenvalues
\be
\lambda=(1-\alpha)\pm \frac{\beta}{\sqrt{3}}i,
\ee
each with degeneracy $2$. Thus, the central fixed point is stable if and only if $(1-\alpha)^2+\frac{\beta^2}{3}<1$. For a fixed choice of $\beta$ one therefore has stability for $\alpha>\alpha_c=1-\sqrt{1-\beta^2/3}$, and an unstable fixed point otherwise.

\begin{figure}
\vspace{3em}
\centerline{\includegraphics[width=0.5\textwidth]{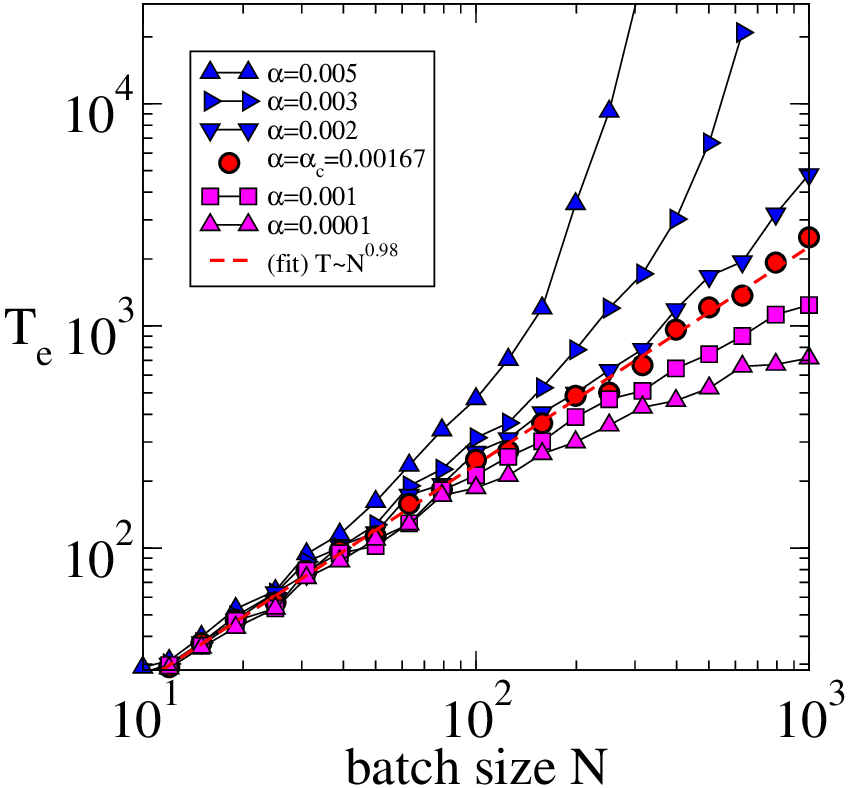}}
\caption{\label{fig:rps} (Colour on-line) Mean escape times in the learning of rock-papers-scissors  at a fixed value of $\beta=0.1$ as a function of the batch size.  Symbols are from numerical simulations, averaged over $100$ independent runs.   The mean escape time is sub-linear in $N$ if the fixed point is unstable ($\alpha<\alpha_c$), and super-linear for $\alpha>\alpha_c$ (stable fixed point). If the central fixed point is neutrally stable ($\alpha=\alpha_c$) the escape time scales linearly in the batch size $N$. The dashed line is a fit to a power law  of the data at $\alpha=\alpha_c$ and reveals a linear scaing $T\sim N^{0.98}$. For the present choice of $\beta=0.1$ one has $\alpha_c\approx 1.668\cdot 10^{-3}$.}
\end{figure}

This separation of two regimes, one with a stable fixed point, and the other with a deterministic flow away from the centre of strategy space, is reflected in the escape times of stochastic learning. Results are shown in Fig. \ref{fig:rps}. In our simulations the stochastic learning dynamics is started at the fixed-point $\bx=(1/3,1/3,1/3), \by=(1/3,1/3,1/3)$ at the centre of the strategy simplex and evolved at finite batch size $N$. The system does not fixate into one pure strategy, so the escape time is measured as the point in time when the $6$-dimensional vector $(\bx,\by)$ first crosses a circle of radius $R=0.1$ around the fixed point. As seen in the figure the escape time scales sublinearly with the batch size if the fixed point is unstable ($\alpha<\alpha_c$). For neutrally stable deterministic dynamics algebraic scaling is found, and escape is sub-extensively slow in the regime of a stable fixed point ($\alpha>\alpha_c$). 
 
\section{Network games}\label{sec:net}

\subsection{Definition of the game}
We will now move to a more complex multi-player game defined on a networked structure, and consider the so-called `best shot game' \cite{galeotti}. Analyses of the statistics of Nash equilibria on random graphs can be found in \cite{asta1,asta2}. We here again focus on adaptive learning.  Players are labelled by $i=1,\dots,M$ and arranged on an undirected graph, so that players $i$ and $j$ interact if and only if the link between $i$ and $j$ is present in the graph. In the  `best shot' game each player has the choice between two actions, to `contribute' or not to contribute. For simplicity we will refer to these actions as $1$ and $0$ respectively. The payoff any given player $i$ receives in any round of the game then depends on her action and on the actions of her neighbours on the underlying network. If we write $\partial i$ for the set of neighbours of player $i$ then the best-shot game is defined by the following payoff structure for action $s_i=1$
\be
u(s_i=1,\bs_{\partial i})=\left\{\begin{array}{cc} a & \mbox{if~} s_j=0~ \forall j\in\partial i \\ 
\\
0 & \mbox{if~} \exists j\in\partial i: s_j=1 \end{array}\right.,
\ee
and by payoffs for action $s_i=0$ 
\be
u(s_i=0,\bs_{\partial i})=\left\{\begin{array}{cc} 0 & \mbox{if~} s_j=0~ \forall j\in\partial i \\ 
\\
b & \mbox{if~} \exists j\in\partial i: s_j=1 \end{array}\right..
\ee
The variables $a$ and $b$ are positive constants. To a certain extent the game resembles the typical structure of public goods games. In absence of any contributors in player $i$'s neighbourhood ($\sum_{j\in\partial i} s_j=0$) player $i$ will increase his payoff by contributing. If however at least one of her neighbours is contributing already   ($\sum_{j\in\partial i} s_j>0$), then player $i$ will not want to contribute herself.

\subsection{Sato-Crutchfield equations and homogeneous fixed point}
We will write $x_{i,s}(t)$ with $i=1,\dots,M$ and $s\in\{0,1\}$ for the probability with which player $i$ takes action $s$ at time $t$. One always has $x_{i,0}(t)=1-x_{i,1}(t)$. In the continuous-time limit one obtains the following deterministic equation
\be
\frac{\dot x_{i,s}}{x_{i,s}} =\beta\sum_{\bs_{\partial i}} \left[u(s,\bs_{\partial i})\prod_{j\in\partial i}x_{j,s_j}\right]-\beta\sum_{s^\prime ,s_{\partial i}}\left[u(s^\prime,\bs_{\partial i})x_{i,s^\prime}\prod_{j\in\partial i}x_{j,s_j}\right]+ \alpha\sum_{s^\prime\in{0,1}}x_{i,s^\prime}\ln\frac{x_{i,s^\prime}}{x_{i,s}}.
\ee
Taking into account that 
\begin{align}
\sum_{s_{\partial i}}u(1,\bs_{\partial i})\prod_{j\in\partial i}x_{j,s_j} &= a\prod_{j\in\partial i}x_{j,0}\\
\sum_{s_{\partial i}}u(0,\bs_{\partial i})\prod_{j\in\partial i}x_{j,s_j} &= b\left[1-\prod_{j\in\partial i}x_{j,0}\right]
\end{align}
we can rewrite the equations above in terms of the parameters $x_i = x_{i,1}$: 
\be\label{e:fpsimple}
\frac{\dot{x}_i}{x_i} =\beta(1-x_i)\left[(a+b) \prod_{j\in\partial i}(1-x_j) -b + r\ln\frac{1-x_i}{x_i}\right]
\ee
where we have introduced $r = \frac{\alpha}{\beta}$.
Up to now all derivations hold for any network structure. In order to keep the analytical expressions to a sensible level we will from now on restrict the analysis to regular graphs, i.e. to graphs in which all players have the same number of neighbours. We will denote the degree of the resulting regular network by $K$. Looking for homogeneous fixed-point solutions of the above continuous-time dynamics, i.e.  setting $x_i(t)\equiv x$ for all players $i$, one finds
\be
\frac{\dot{x}}{x}=\beta (1-x)\left[(a+b) (1-x)^{K}-b+r\ln\frac{1-x}{x}\right].
\ee 
Excluding trivial fixed points, i.e. assuming $x\neq0$ and $x\neq 1$, one obtains 
\be\label{e:fp}
(a+b)(1-x)^K+r\ln\frac{1-x}{x} =b 
\ee

The solutions to this equation give the possible fixed points of the deterministic learning dynamics. In the cases studied here we will typically have one internal fixed point, $x^\ast$, its numerical value will generally depend on the model parameters $r,a$ and $b$.  For $a=b$ and $r =0$ we recover the homogeneous mixed Nash equilibrium $x^\ast_0 =1-1/2^{1/K}$. Irrespective of $a$ and $b$ one finds $x^*\approx 1/2$ for $r\gg 1$.

\subsection{Stability analysis}
Expanding Eq. (\ref{e:fpsimple}) around the fixed point $x_i=x^\ast+\widetilde x_i$ to linear order one finds

\be\label{e:jacobian}
\dot{\widetilde x}_i =-\beta \left(\rho\sum_{j\in\partial i}\widetilde x_j + r\widetilde x_i\right), 
\ee
where $\rho = (a+b) x^\ast(1-x^\ast)^K$. Eq. (\ref{e:jacobian}) can then be written in matrix form as 
\be
\dot{\bf \widetilde x} = -\beta \left(\rho \mathbb{A}+r \mathbb{I}\right){\bf \widetilde x},
\ee
where $\mathbb{A}$ is the adjacency matrix of the graph and $\mathbb{I}$ the $M\times M$ identity matrix. Diagonalizing this equation is equivalent to diagonalizing $\mathbb{A}$. In particular the critical value of $r$, separating the phase $r>r_c$ in which the fixed point $x^\ast$ is stable from a phase with an unstable fixed point $(r<r_c)$ is given by
\be\label{eq:rc}
r_c=-\mu \rho,
\ee
where $\mu$ is the smallest eigenvalue of the adjacency matrix $\mathbb{A}$ (all eigenvalues are real since $\mathbb{A}$ is symmetric).
 
 \begin{figure}
\vspace{3em}
\centerline{\includegraphics[width=0.5\textwidth]{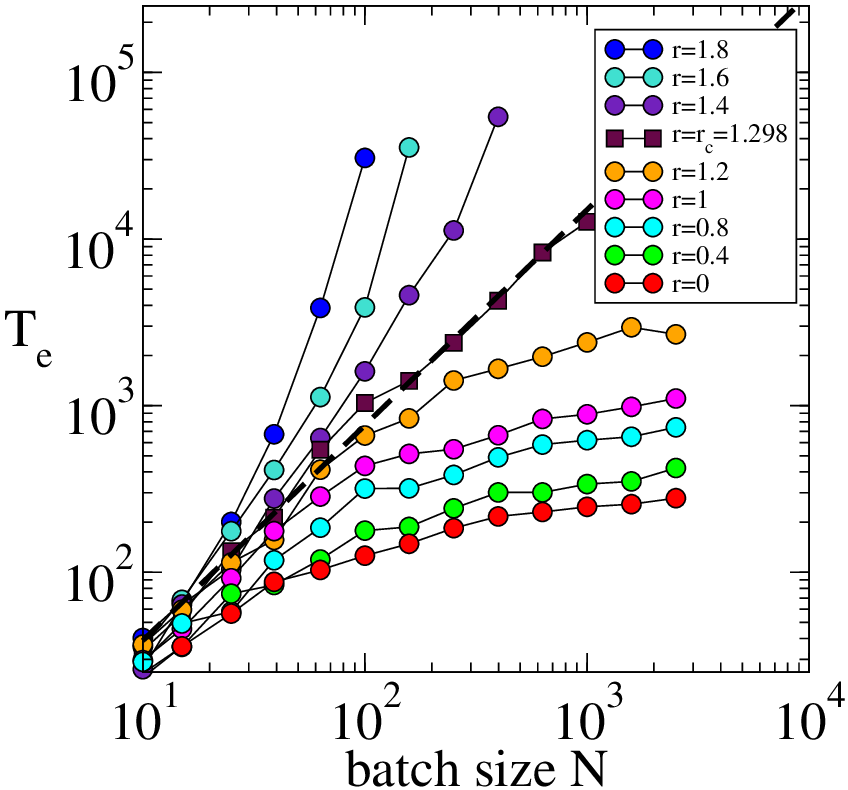}}
\caption{\label{fig:networkescape}(Colour on-line) Escape times of learning in the networked best-shot game ($a=7,b=1$, $\beta=0.01$). Simulations are performed on a fixed regular graph with $M=10$ players and $K=3$, created at random. All runs are started at the homogeneous fixed point $x_i=1/2$, the escape time is defined as the first time in each run when $\sum_{i=1}^{10} (x_i(t)-1/2)^2=10^{-3}$. Results are averaged over $20$ independent runs (all on the same realisation of the graph). The most negative eigenvalue of the adjacency matrix of this specimen graph is approximately $-2.596$, so that Eq. (\ref{eq:rc}) predicts $r_c\approx 1.298$. An exponent of approximately $1.3$ is found in a power law fit to the escape time at $r=r_c$ (dashed line).}
\end{figure} 

In analogy with the earlier sections we expect that the escape time of learning will scale logarithmically in the batch size for $r<r_c$ when the interior fixed point is unstable. In the stable phase ($r>r_c$) on the other hand one would predict an exponential behaviour. We verify these predictions in the following section. As a final remark regarding stability it is interesting to consider the limiting case of deterministic learning started from homogeneous initial conditions $x_i(t=0)=x(0)$ for all $i$. For regular networks of degree $K$ one then has $x_i(t)=x(t)$ for all $i$, and $x(t)$ fulfills

\be\label{e:fpsimple2}
\dot{x} =\beta x(1-x)\left[(a+b) (1-x)^K -b + r\ln\frac{1-x}{x}\right].
\ee
Linearising about the fixed point, and restricting the motion to the space $\{x_i\equiv x\}$, one finds
\be
\dot{\widetilde x} = -\beta \left(\rho K+r\right){\widetilde x}.
\ee
Given that $\rho K>0$ the interior fixed point is therefore stable irrespective of the value of the parameter $r\geq 0$, similar to what was observed in the Hawk-Dove game. The network game considered in this section therefore bears close similarity to the Hawk-Dove game discussed earlier on. In a one-population setting (equivalently upon restricting the dynamics to the subspace $x_i(t)\equiv x(t)$) the deterministic dynamics has a stable internal fixed point for any $r\geq 0$. In the multi-population case the fixed point remains unchanged, but unstable eigenvalues are present for $r<r_c$. The corresponding eigendirections break the symmetry between the different co-ordinates $\{x_i\}$, and hence the flow is away from the manifold defined by $x_i(t)\equiv x(t)$.

 \subsection{Test against simulations}
The above theoretical predictions can be tested in several possible ways. For example one can consider the thermodynamic limit of large regular random networks of degree $K$, and then perform an average over multiple instances of the graph. Using results from spectral graph theory \cite{mckay, cioaba} the support of the eigenvalue distribution of the adjacency matrix of a large regular random graph of degree K typically has its most negative eigenvalue at
\be
\mu=-2 \sqrt{K-1}.
\ee
With this estimate the expected value of $r_c$ can then be computed by means of Eq. (\ref{eq:rc}). Simulations of the learning dynamics on large networks are however time consuming, and we have therefore taken a different route. We have created one particular instance of a regular random graph with $M=10$ nodes and degree $K=3$. The adjacency matrix of this particular graph has then been diagonalised and the relevant eigenvalue has been identified as $\mu\approx -2.596$. For convenience we have also chosen $a=7, b=1$, ensuring the value $x^*=1/2$ for the deterministic fixed point\footnote{This choice was made purely for analytical convenience, to ensure that the deterministic fixed point carries no dependence on $r$. If one were to use the best-shot game to model an actual public goods game one would probably choose $b>a$.}. Eq. (\ref{eq:rc}) then predicts a change of stability at $r_c\approx 1.298$. Measurements of the escape time of the best-shot game on this fixed sample of the graph are shown in Fig. \ref{fig:networkescape}. Results are consistent with an algebraic dependence of the escape time $T_e$ on the batch size at $r=r_c$, even though we note a slight discrepancy from the exponent of unity one would expect from the Langevin approximation\footnote{We attribute these differences to the approximation implied by the continuous-time limit, and to potential effects of higher-order terms in the expansion in powers of $N^{-1/2}$.}. Below $r_c$ the fixed point is unstable, and escape times are shorter, consistent with logarithmic scaling in the batch size. At $r>r_c$ the fixed point is stable and escape is slow. 

\begin{figure}
\vspace{3em}
\centerline{\includegraphics[width=0.5\textwidth]{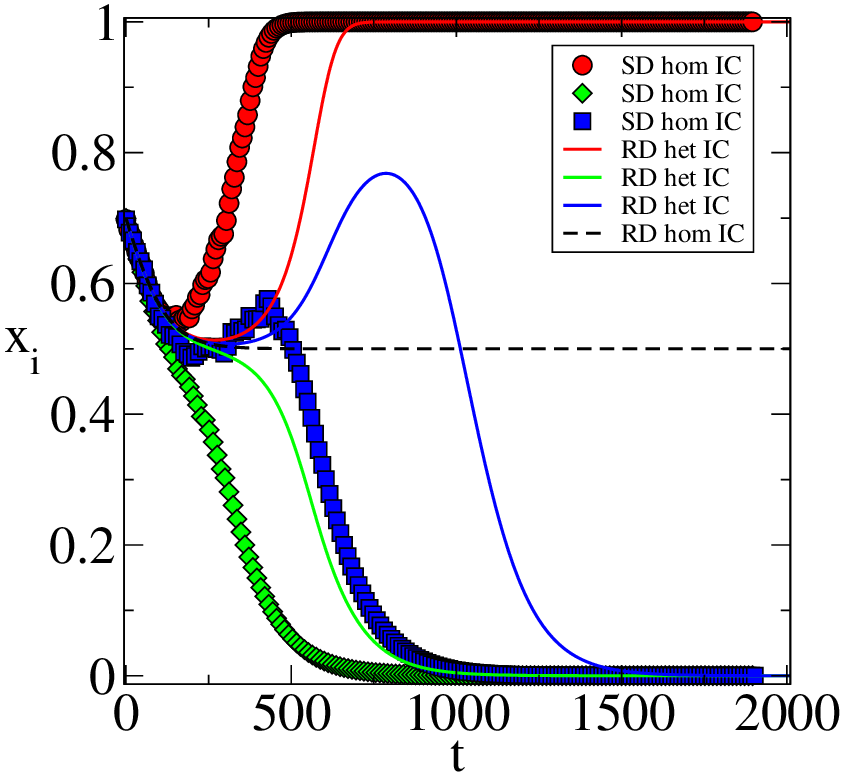}}
\caption{\label{fig:networkfixation} (Colour on-line) Fixation in learning in the networked best-shot game ($a=7, b=1$). Simulations are performed on a fixed regular graph with $M=10$ players and $K=3$, created at random. The black dashed line indicates the evolution of the deterministic replicator dynamics (RD), started with homogeneous initial conditions (hom IC) $x_i=0.7$ for all players $i$. Symbols denote the evolution of the corresponding stochastic learning dynamics (SD) with batch size $N=10$; for clarity, only three players are followed but all of them fixate. The colour lines show the evolution, for the same three players, of the deterministic replicator equations when the homogeneous initial conditions above are slightly perturbed in the direction of the configuration to which the stochastic learning fixated (het IC). That is to say, $x_i=0.7+\delta_i$ where $\delta_i$ is a random number smaller than $10^{-3}$ which is positive if the configuration reached by the stochastic learning has $x_i = 1$ and it is negative otherwise. If the perturbation is in any other direction, the configuration to which the replicator equations fixate may be different from the one reached by stochastic learning.}
\end{figure}

Fig. \ref{fig:networkfixation} illustrates the dynamical evolution under either the deterministic replicator equations or the stochastic learning dynamics $(r=0)$. It shows how the system is driven to fixation by a non-homogeneous perturbation, caused either by a slight heterogeneity in the initial conditions of the replicator equations or by the stochasticity induced by a finite batch size in the learning dynamics. Specifically one finds that (i) the fixed point $x_i=x^*$ is an attractor for homogeneous initial conditions\footnote{We expect that this no longer holds on graphs which are not regular, i.e. in which different nodes have different degrees. Such a heterogeneity can be expected to break the symmetry between the nodes.} and that (ii) inhomogeneous initial conditions lead to a flow towards the corners of strategy space.

\section{Conclusions}
In summary we have studied the fixation properties of simple reinforcement-type learning algorithms in the context of different games, and have compared them to the outcome of evolutionary dynamics.The examples we have chosen range from simple two-player two-action games to more complex multi-player games on networked structures. Our main results can be summarized as follows: (i) Unlike evolutionary dynamics in which fixation can occur purely driven by fluctuations, fixation (i.e. convergence to pure strategies) in learning of games of the type we have studied here appears to be possible only if the underlying deterministic dynamics itself converges to a pure action profile. This is typically only the case if the symmetry between players is broken, for example by an inhomogeneous initial condition; (ii) Two-player and multi-player learning in the deterministic limit can, to a good approximation, be described by equations of a multi-population replicator type. As seen for the Hawk-Dove game, and for the network game we have studied, the stability properties of multi-population replicator dynamics can differ substantially from those of the corresponding one-population model; (iii) The role of noise in fixation processes in dynamical learning is mostly limited to triggering the required breaking of symmetry, eventually leading to fixation, unlike in evolutionary processes we have not found examples in which fixation is triggered by random drift alone. (iv) In cases for which the limiting deterministic learning converges to a symmetric fixed point in the interior of strategy space the corresponding escape time depends on the stability of this fixed point: for stable fixed points escape is essentially exponential, for unstable fixed points logarithmic scaling in the batch size is found, these findings are very similar to those found in evolutionary systems.

While we have pointed out crucial differences between multi-player learning and evolutionary dynamics, our results mostly extend the similarities between the two approaches to dynamical aspects of games. In \cite{galla} it was pointed out that stochastic learning can exhibit persistent quasi-cycles in regimes where deterministic learning converges to fixed points. These effects are very similar to those observed in evolutionary systems. The present work shows that the analogy goes further, and that the escape and fixation properties of stochastic learning dynamics are closely related to the corresponding behaviour of population models. Our analysis shows that the analogy is particularly strong when learning is compared with evolution in multiple populations. We expect that these similarities stretch even further, including potentially pattern formation in spatially extended systems and or more complicated dynamics on adaptive networks. Future work may also include more complex learning models \cite{camerer1, Camerer2003, Ho2007} inspired by laboratory experiments in behavioural game theory or by algorithms in machine learning. We are here confident that the analogy with stochastic evolutionary systems may provide a powerful perspective and that it can contribute to accelerating the research required to analyze more general learning dynamics.


\begin{acknowledgments} 
TG and JRG acknowledge hospitality of the Abdus Salam ICTP. TG is grateful for funding by the Research Councils UK (RCUK reference EP/E500048/1) and by EPSRC (references EP/I005765/1 and EP/I019200/1), and would like to thank A. J. Bray for useful discussions. BSZ acknowledges support by an EPSRC studentship.

\end{acknowledgments}

\section*{Appendix: Escape time for linear Langevin dynamics}
\subsection{Reduced problem and backward Fokker-Planck equation}
 Let us consider the simple 1D problem
\be\label{eq:1}
\dot x = \lambda x + \sqrt{2D}\eta
\ee
where $\eta(t)$ is standard Gaussian white noise, i.e. $\avg{\eta(t)}=0$ and
\be\label{eq:eta}
\avg{\eta(t)\eta(t')}=\delta(t-t').
\ee
We consider escape times: Fix a number  $R>0$, if the process is started at $t=0$ at $x(0)=0$, then the escape time is defined as the first time, the process leaves the interval $[-R,R]$.

Using standard methods \cite{gardiner, risken} one finds the backward Fokker-Planck equation
\be
\lambda x g'(x)+D g''(x)=-1,
\ee
where $g(x)$ is the mean escape time, conditioned on a starting point $x$. Boundary conditions read $g(R)=g(-R)=0$ (if the process is started at the boundaries of the interval, the extinction time is trivially zero). It is easy to check that the solution is given by
\be\label{eq:4}
g(x)=\frac{1}{D}\int_x^R~ dy ~e^{-\lambda y^2/(2D)}\int_0^y ~dz~ e^{\lambda z^2/(2D)}.
\ee
Restricting to a starting point of $x=0$ one has
\be\label{eq:5}
g(0)=\frac{1}{D}\int_0^R~ dy ~e^{-\lambda y^2/(2D)}\int_0^y ~dz~ e^{\lambda z^2/(2D)}.
\ee

By introducing the change of variables $z=R \sqrt{(1-u)v}$ and $y= R\sqrt{v}$, introducing the appropriate Jacobian $\frac{R^2}{4\sqrt{1-u}}$, and setting $w_N=\lambda R^2/2D$, we can rewrite the integrals on the RHS of Eq. (\ref{eq:5}) as 
\be
g(0)= \frac{w_N}{\lambda}\mathop{{{}_{{2}}F_{{2}}}\/}\nolimits\!\left({1,1\atop 2,\frac{3}{2}};-w_N\right),
\ee
where 
\be\label{eq:ghf}
\mathop{{{}_{{2}}F_{{2}}}\/}\nolimits\!\left({1,1\atop 2,\frac{3}{2}};z\right)=\frac{1}{2}\int_0^1 \mathrm{d}v ~\int_0^1\mathrm{d} u~\frac{e^{z u v}}{\sqrt{1-u}} \ ,
\ee
is a generalised hypergeometric function whose asymptotic behaviour is well known \cite{DLMF}. For simplicity, we will in the following give a heuristic derivation of the asymptotic leading behaviour of $g(0)$ at large values of $N$. Our starting point will be the expression given in Eq. (\ref{eq:5}). 

\subsection{Large-$N$ behaviour for $\lambda>0$}
Setting $2D=\sigma^2/N$, so that $w_N = N\lambda R^2/\sigma^2$, and changing variables in (\ref{eq:5}) to $u=z+y$ and $v=z-y$ one has
\be
g(0)=\frac{N}{\sigma^2}\int_{-R}^0\mathrm{d} v\int_{-v}^{v+2 R}\mathrm{d}u ~ e^{N\lambda u v/\sigma^2}.
\ee
Executing the integration over $u$ and setting $w=N\lambda R v/\sigma^2$ one finds
\be\label{eq:5b}
g(0)=\int_\delta^{w_N} \mathrm{d}w ~ \frac{e^{-w^2/w_N}}{\lambda w}-\int_\delta^{w_N} \mathrm{d}w ~\frac{ e^{-2 w + w^2/w_N}}{\lambda w},
\ee
where we have introduced the lower integration limit $\delta$. This variable will be set to zero eventually, the purpose of our procedure is to keep track off singularities at small values of $w$. 

When $\lambda>0$ and $N$ is very large the main contribution to the integral in (\ref{eq:5b}) comes from small values of $w$ due to the decaying exponential factors. Notice that the term $w^2/w_N$ in the second integral in (\ref{eq:5b}) is only relevant when $w\approx w_N$ in which case the exponent $-2w+w^2/w_N \approx -w_N$ and so the integrand is exponentially small in $w_N$. Therefore, we can neglect the term $w^2/w_N$ in the exponent relative to $w$. Doing this and introducing the variable $\Omega = w^2/w_N$ in the first integral and $\Omega^\prime = 2w$ in the second, we obtain
\be\label{eq:5c}
g(0)\approx \frac{1}{2\lambda}\int_{\delta^2/w_N}^{w_N} \mathrm{d}\Omega ~\frac{e^{-\Omega}}{\Omega}-\frac{1}{\lambda}\int_{2 \delta}^{2 w_N} \mathrm{d}\Omega^\prime ~\frac{e^{-\Omega^\prime}}{\Omega^\prime}.
\ee
Both integrals on the RHS are of the type $\int_\epsilon^{x}\mathrm{d}\Omega ~\frac{e^{-\Omega}}{\Omega}$, where $\epsilon=\delta^2/w_n$ in the first case, and $\epsilon=2\delta$ in the second. In both cases $\epsilon$ tends to zero as $\delta\to 0$.  The upper integration limit $x$ is proportional to $w_N$ in both integrals. Integrals of this type can be simplied by an integration by parts, and we find
\be\label{eq:5d}
\int_\epsilon^{x}\mathrm{d}\Omega ~\frac{e^{-\Omega}}{\Omega}=\int_\epsilon^\infty\mathrm{d}\Omega ~\frac{e^{-\Omega}}{\Omega}-\int_x^\infty\mathrm{d}\Omega ~\frac{e^{-\Omega}}{\Omega}=\left. e^{-\Omega}\log\Omega\right|_\epsilon^\infty + \int_\epsilon^\infty\mathrm{d}\Omega ~ e^{-\Omega}\log\Omega -\int_{x}^\infty\mathrm{d}\Omega \frac{e^{-\Omega}}{\Omega}.
\ee
The last integral in (\ref{eq:5d}) is exponentially small in $x\propto w_N$, as one can see from
\be
\int_{x}^\infty\mathrm{d}\Omega ~\frac{e^{-\Omega}}{\Omega}\leq \int_{x}^\infty\mathrm{d}\Omega ~e^{-\Omega}=e^{-x},
\ee
valid for $x\geq 1$. We can therefore neglect the last term in Eq. (\ref{eq:5d}) in the limit of $N\to\infty$, when we have $w_N\to\infty$.
Performing the limit $\delta\to 0$ (and taking into account that $\epsilon\to 0$ in this limit) we find
\be\label{eq:5e}
\int_\epsilon^{x}\mathrm{d}\Omega ~\frac{e^{-\Omega}}{\Omega}\approx -\log\epsilon - \gamma_e ,
\ee
where $\gamma_e = -\int_0^\infty e^{-\Omega}\log\Omega = 0.57721\ldots $ is the Euler-Mascheroni constant.  Using these results in Eq. (\ref{eq:5c}), we finally find
\be\label{eq:lgtz}
g(0)\approx \frac{1}{2\lambda}\left(\log w_N + \gamma_e + \log 4\right)
\ee
Except for the additional $\log 4$ term, this result coincides with the asymptotic results for escape times reported in [Mobilia 2010] for the case of a two-dimensional system.

\subsection{Large-$N$ behaviour for $\lambda<0$}
For $\lambda<0$ and $N$ large, we have from Eq. (\ref{eq:5})
\be\label{eq:9}
g(0)=\frac{2N}{\sigma^2}\int_0^R~ dy \int_0^y ~\mathrm{d}z ~ e^{N|\lambda | (y^2-z^2)/\sigma^2}\approx \frac{2N}{\sigma^2}\int_0^R~ \mathrm{d}y ~ e^{N|\lambda | y^2/\sigma^2} ~ \frac{1}{2}\int_{-\infty}^\infty ~\mathrm{d}z~  e^{-N|\lambda| z^2/\sigma^2}.
\ee
We have here first extended the integration range of the $z$-integration to the interval $[-y,y]$, adjusted for by an overall factor of $1/2$. Subsequently, based on the observation that the integrand assumes its maximum at $z=0$ we have extended the integration range of the $z$-integral to the entire real axis. This introduces an error which does not contribute to the leading exponential behaviour for large $N$. The integral over $z$ is now Gaussian, and can be evaluated straightforwardly. The exponent in the remaining $y$-integration reaches its maximum at $y=R$. Expanding the exponent up to first order about this value we get
\be\label{eq:lltz}
g(0)\approx \frac{1}{2}\sqrt{\frac{\pi\sigma^2}{N|\lambda|}}\frac{2N}{\sigma^2}~e^{-N|\lambda |R^2/\sigma^2}\int_0^R~ \mathrm{d}y ~ e^{2 N |\lambda | R y/\sigma^2}\approx\frac{1}{2|\lambda|}\sqrt{\frac{\pi}{|w_N|}} ~{e^{|w_N|}},
\ee
with $|w_N| = N|\lambda|R^2/\sigma^2$.

 \end{document}